# CLUSTERING WEB SEARCH RESULTS FOR EFFECTIVE ARABIC LANGUAGE BROWSING

## Issam SAHMOUDI[1] and Abdelmonaime LACHKAR[1]

[1]LSIS, ENSA - Sidi Mohamed Ben Abdellah University (USMBA), FEZ, Morocco

`Issam.sah@gmail.com, abdelmonaime_lachkar@yahoo.fr`

## ABSTRACT

*The process of browsing Search Results is one of the major problems with traditional Web search engines for English, European, and any other languages generally, and for Arabic Language particularly. This process is absolutely time consuming and the browsing style seems to be unattractive. Organizing Web search results into clusters facilitates users quick browsing through search results. Traditional clustering techniques (data-centric clustering algorithms) are inadequate since they don't generate clusters with highly readable names or cluster labels. To solve this problem, Description-centric algorithms such as Suffix Tree Clustering (STC) algorithm have been introduced and used successfully and extensively with different adapted versions for English, European, and Chinese Languages. However, till the day of writing this paper, in our knowledge, STC algorithm has been never applied for Arabic Web Snippets Search Results Clustering.*

*In this paper, we propose first, to study how STC can be applied for Arabic Language? We then illustrate by example that is impossible to apply STC after Arabic Snippets pre-processing (stem or root extraction) because the Merging process yields many redundant clusters. Secondly, to overcome this problem, we propose to integrate STC in a new scheme taking into a count the Arabic language properties in order to get the web more and more adapted to Arabic users. The proposed approach automatically clusters the web search results into high quality, and high significant clusters labels. The obtained clusters not only are coherent, but also can convey the contents to the users concisely and accurately. Therefore the Arabic users can decide at a glance whether the contents of a cluster are of interest. Preliminary experiments and evaluations are conducted and the experimental results show that the proposed approach is effective and promising to facilitate Arabic users quick browsing through Search Results. Finally, a recommended platform for Arabic Web Search Results Clustering is established based on Google search engine API.*

## 1. INTRODUCTION

Web search engines have become indispensable in finding relevant information from the World Wide Web. The Number of Arabic documents available in the Internet is growing enormously. It has become more and more interesting for Arabic users to find relevant Arabic documents. In response to the users using Arabic query, the available search engines such as (Google: http://www.google.com, Yahoo: http://www.yahoo.com and Bing: http://www.bing.com ) return a ranked list of search results (**Snippets**) that contain a mixture of documents from Arabic Language and other languages to response to Arabic Query. Even if, we can personalize the user query to make the results just in Arabic Language the ranked list is highly inefficient since the number of retrieved search results can be in the thousands for a typical query.

Most users just view the top results displayed in the first pages and therefore might miss relevant Arabic documents. In such a case, it would be better to filter and to group the search results into particular interesting groups, so clustering view of the search results would be much more useful to a user than a simple ranked list.

Many different solution and realization have been presented in the last years (Clusty: http:// www.clusty.com, iBoogie: http:// www.iboogie.com, yippy: http:// www.yippy.com) [1]. These solutions have been developed especially for Latin Language or use Cross-language from Arabic Language to English Language to construct different clusters using different clustering algorithms. In such case the search results clusters corresponding to the user query contain also a mixture of documents from different languages.

Note that the **Data-Centric Clustering Algorithms** such as K-mean and Agglomerative Hierarchical Clustering algorithms are inadequate, since they don't generate clusters with highly readable names or cluster labels. To solve this problem, **Description-Centric Algorithms** such as Suffix Tree Clustering (STC) algorithm have been introduced and used successfully and extensively with different adapted versions for English [2], European [2], and recently for Chinese Languages [3][4]. However, till the day of writing this paper, in our knowledge STC algorithm has been never applied to Arabic web **Snippets** search results clustering. In this paper, we propose first, to study how STC can be applied to Arabic Language? We then illustrate by example that is impossible to apply STC after **Snippets** pre-processing (stem or root extraction) because the Merging process yields many redundant clusters. Secondly, to overcome this problem, we propose to integrate STC in a new scheme taking into a count the Arabic language properties in order to get the web more and more adapted to Arabic users.

The remainder of this paper is organized as follows. In the next section, we begin by discussing the related works. Section 3 presents the Arabic Language properties, the STC algorithm, and illustrates the related problem. To overcome this problem, in the section 4 we propose to integrate STC in a new scheme in order to get the web more and more adapted to Arabic Language. Preliminary experiments and evaluations are conducted in section 5. Finally, we provide conclusion and future works in section 6.

## 2. RELATED WORKS

The clustering of web search results is to organize the **Snippets** sharing a common topic into the same clusters, and form the corresponding labels for description. It plays an important part in modern search engine. Related studies have been developed in recent years. The first dynamic system is the Scatter-Gather system which is particularly helpful in the queries difficult to specify formally [5]. Scatter-Gather system is followed by another on-line search results clustered system named Grouper dynamically grouping the **Snippets** into clusters interpreted by the extracted phrases [2]. In this section, we'll briefly survey, discuss, and compare the search results clustering techniques that have been proposed to group into clusters the Ranked list returned as search results of any Web Search Engine such as Google, Yahoo, and Bing.

These latter, may be categorized into two clustering approaches: **Data-Centric** and **Description-Centric**. In the **Data-Centric** approach, we found conventional data clustering algorithms (e.g., hierarchical agglomerative, k-means, spectral) applied to search results and often slightly adapted to produce a comprehensible cluster description: Input texts are represented as Bags of Words, and cluster labels typically consist of single words recovered from the cluster representative, possibly expanded into phrases. This class contains both early and recent systems, such as Lassi[6], CIIRarchies[7], Armil[8], with Scatter/Gather[5] as remarkable predecessor.

Note that the main drawback of this approach is related to the quality of cluster labels which is the most critical part of any Web Search Results Clustering. To overcome this problem when the quality of cluster labels is given priority of search results, we can speak about the **Description-Centric approach**. This approach, pioneered by Suffix Tree Clustering (STC) [2] has become main stream in the last few years [9].The  STC is a on-line clustering technique where search results (mainly **Snippets**) are clustered fast (in linear time), incrementally, and each cluster is labelled with a common phrase. Another advantage of STC is that it allows clusters to overlap. That's why it is the most algorithms have been used or ameliorated for web search results clustering. But the STC also have some inconvenient it can't be applied directly to any language because each one has their own characteristic, such as Chinese Language. There's many works to adapt STC and improving search web results clustering for Chinese [10]. As Chinese Language, Arabic also has different morphology than Latin Language. Therefore, using STC directly for Arabic language can influence negatively the quality of the search results clusters.  In the following sections, we propose to study and propose some improvement and adaptation of the STC Algorithm for Arabic Language.

## 3. SUFFIX TREE CLUSTERING FOR ARABIC LANGUAGE:

The Arabic web presents an important portion of the web. With Arabic as the 5th most spoken language in the world and with the increasing number of Arabic internet users and the increasing number of Arabic documents available in the internet at exponential rates, it has become more and more interesting for Arabic users to find relevant Arabic documents [11]. The Arabic Language morphology is so different than Latin Language [12].

In our knowledge there is no study or proposition has been done to improve Arabic search results using **Suffix Tree Clustering** algorithm.  In the following, we present the Arabic Language properties, the Suffix Tree Clustering algorithm, and we illustrate by example the related problem.

### 3.1. Arabic Language Proprieties:

Arabic is one of the oldest languages in the world. It belongs to the Semitic family of languages. It is significantly different from English and other European languages in a number of important respects [13]:

- It is written from right to left.
- It is mainly a consonantal language in its written forms, i.e. it excludes vowels.
- Its two main parts of speech are the verb and the noun in that word order, and these consist, for the main part, of trilateral roots (three consonants forming the basis of noun forms that derive from them).
- It is a morphologically complex language in that it provides flexibility in word formation: complex rules govern the creation of morphological variations, making it possible to form hundreds of words from one root.

Arabic consists of two parts: the consonants (letters) and the vowel signs (over and underscores used with letters to indicate proper pronunciation). The consonants are the more important part of the written language, for they convey the basic meanings of the words. In fact, the common practice in written Arabic is to omit vowel signs in all published works, except poetry and some religious texts. The following is a list of other characteristics that could create potential problems for Arabic **Information Retrieval** (IR) that don't apply to English IR [13]:

- Some Arabic words, particularly the definite article ' , al' and a number of conjunctions and prepositions, are not separated from their following word by a space. This results in a large number of entries being clustered together alphabetically in index files.
- The peculiar morphology of Arabic might render methods used for English text retrieval inappropriate. As an example, the English phrase "and she wrote it"comprising four words would be written in Arabic as one word"ه    " ( =and,    =wrote,  =she, =it). In this case " " (meaning "and") has been slightly transformed and linked and with the following word; this would create problems were it decided to treat " " as a stopword.
- It is common to find many Arabic words that have different pronunciations and meanings but share the same written form (homonyms), making finding the appropriate semantic occurrence of a given word a problem. English also has many homonyms, of course, but the problem is aggravated in Arabic by the absence of vowels in the written form, which then produces many identical consonant groupings.
- Unlike English, truncating the beginning or end of an Arabic word does not lead invariably to its root.
- Arabic plurals are formed more irregularly than in English; depending on the root and the singular form of the word, the plural form might be produced by the addition of suffixes, prefixes or infixes, or by a complete reformulation of the word
- A double letter in Arabic is denoted with a pronunciation mark (shadda).

## 3.2 SUFFIX TREE CLUSTERING:

A **Suffix Tree** is a data structure that allows efficient string matching and querying. It have been studied and used extensively, and have been applied to fundamental string problems such as finding the longest repeated substring, strings comparisons , and text compression[14]. The Suffix Tree commonly deals with strings as sequences of characters, or with documents as sequences of words. A **Suffix Tree** of a string is simply a compact tree of all the suffixes of that string. The **Suffix Tree** has been used firstly by Zamir et al. [2] as clustering algorithm named **Suffix Tree Clustering** (STC). It's linear time clustering algorithm that is based on identifying the shared phrases that are common to some **Document's Snippets** in order to group them in one cluster. A **phrase** in our context is an ordered sequence of one or more words. The **Base Cluster** is a set of results that share a common phrase. STC has three logical steps:

- Document's **Snippets** "Cleaning",
- Identifying **Base Clusters** using a Suffix Tree data structure,
- Mergging Base Clusters to generate **Base Clusters Graph.**

## 3.2.1 Document's Snippets "Cleaning"

In this step, each snippet is processed for Arabic stop-words removal such as (e.g.,  وهذا          فانه لهذا): Stop-word means high frequency and low discrimination and should be filtered out in the IR system. They are functional, general, and common words of the language that usually do not contribute to the semantics of the documents and have no read added value. Many Information retrieval systems (IRS) attempt to exclude stop-words from the list of features to reduce feature space, and to increase their performance. In addition, in our case, to deal especially with Arabic snippets, we propose also in this step to remove Latin words and specials characters such as (e.g. $, #, /, - …).

### 3.2.2 Identifying Base Clusters using a Suffix Tree data structure:

The main advantage of STC over methods utilizing term frequency distribution only is that phrases are usually more informative than unorganized set of keywords, and can be directly used to label the discovered clusters, which in other clustering algorithms becomes a problem. This method treats documents as a set of phrases (sentences) not just as a set of words. The sentence has a specific, semantic meaning (words in the sentence are ordered). Suffix tree document model considers a document d = $w_1w_2...w_m$ as a string consisting of words $w_i$ , not characters (i = 1; 2;...; m). A revised definition of suffix tree is follow:

A **Generalized Suffix Tree** for a set S of n strings, each of length $m_n$, is a rooted directed tree with exactly $m_n$ leaves marked by a two number index (k,*l*) where k ranges from 1 to n and *l* ranges from 1 to $m_k$. Each internal node, other than the root, has at least two children and each edge is labeled with a nonempty substring of words of a string in S. No two edges out of a node can have edge labels beginning with the same word. For any leaf (i,j), the concatenation of the edge labels on the path from the root to leaf(i, j) exactly spells out the suffix of $S_i$ that starts at position j , that's it spells out $S_i$[ j...$m_i$ ][15].The figure.1 shows an example of the generated Suffix Tree of a set of three Arabic strings or three Documents–Document1: "**الجبن ياكل** " , Document2: **فارياكل** "**الجبن ايضا**", Document3: "**القط ياكل الفارايضا** ", recursively the figure.2 shows the same example in English Language (Document1: "**cat ate cheese**", Document2: "**mouse ate cheese too** " and Document3: "**cat ate mouse too**") .

In the **Suffix Tree** data structure, every node contains the following information: the current substring and IDs of the **Snippets** that the phrase belongs to. Every **node** of the **Suffix Tree** represents a cluster, which is referred to as a **Base Cluster**. All the Snippets in the same cluster share the phrase it stands for. Each **Base Cluster** should have a score S(B). The clusters can be ranked according to its score. The score is relevant to:

    1) The length of the common words.
    2) The number of the documents in the cluster.

$$S\ (B) = |B|*F\ (|P|)$$

$$F(P) = \begin{cases} 0, if\ |p| = 1 \\ |\ p|, if\ 6 \geq\ |p| \geq 2 \\ \alpha\ ,\ if\ |p| > \end{cases}$$

Where |B| is the number of documents in B, and |P| is the number of words making up the phrase P, and is a constant. Then all base clusters are sorted by the scores, and the top k base clusters are selected for cluster merging in Step 3.

### 3.2.3 Merging Base Clusters to generate Base Clusters Graph:

We define a binary similarity measure between **Base Clusters** based on the overlap of their document sets. Given two **Base Clusters** B1 and B2, with sizes | B1| and | B2| respectively, and *|B1 U B2|* representing the number of documents common to both phrase clusters, we define the similarity **Similarity(B1, B2 )** of B1 and B2 to be :

$$\text{Similarity (B1, B2)} = \begin{cases} 1, if \left(\frac{|B2 \cup B1|}{|B1|}\right) \geq \alpha \; and \; \left(\frac{|B2 \cup B1|}{|B2|}\right) \geq \alpha \\ \\ 0, Otherwise \end{cases}$$

Where   is a constant between 0 and 1

In our system, we set   to be 0.6. Two nodes with similarity of 1 are connected and a **Base Cluster Graph** is constructed. All the connected based clusters are combined into a single final  cluster,  which contains the union of the documents of all the component **Base Clusters** Figure.3 ,Figure.4 .

**Arabic Documents:**

Document (1,"يأكل الجبن          ")                                    A

Document (2, ايضا فارياكل     ")

Document (3," القط ياكل الفارايضا   ")

```
STDM {                                                           B
      1 -> [label="الفعلايكل" doc:(1,3)];
            2 -> [label="ابض $" doc:(1)];
            2 -> [label="الفارايضا $" doc:(3)];
      1 -> [label="يكل" doc:(1,2,3)];
            2 -> [label="ابض" doc:(1,2)];
                  3 -> [label="$" doc:(1)];
                  3 -> [label="ابضا $" doc:(2)];
            2 -> [label="الفارايضا $" doc:(3)];
      1 -> [label="ابض" doc:(1,2)];
            2 -> [label="$" doc:(1)];
            2 -> [label="ابضا $" doc:(2)];
      1 -> [label="$" doc:(1,2,3)];
      1 -> [label="الفار" doc:(2,3)];
            2 -> [label="ياكل الجبن ايضا" doc:(2)];
            2 -> [label="ابضا $" doc:(3)];
      1 -> [label="ابضا $" doc:(2,3)];
}
```

Output of our Java Programme

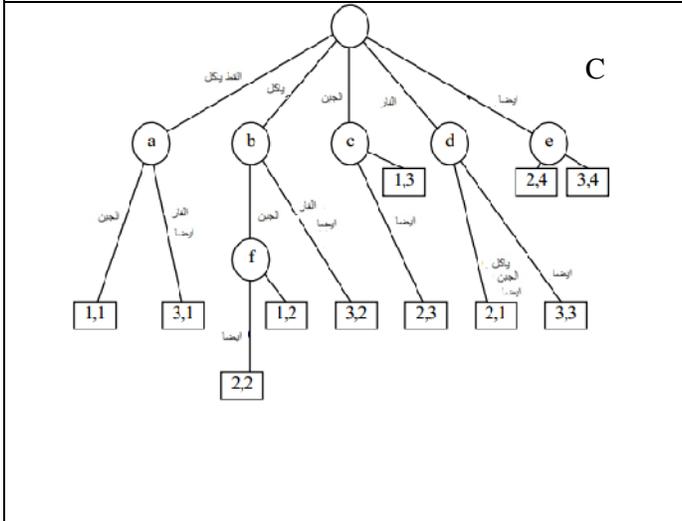

C

Figure.1. An instance of the Suffix Tree using **Arabic Language** (A,B,C)

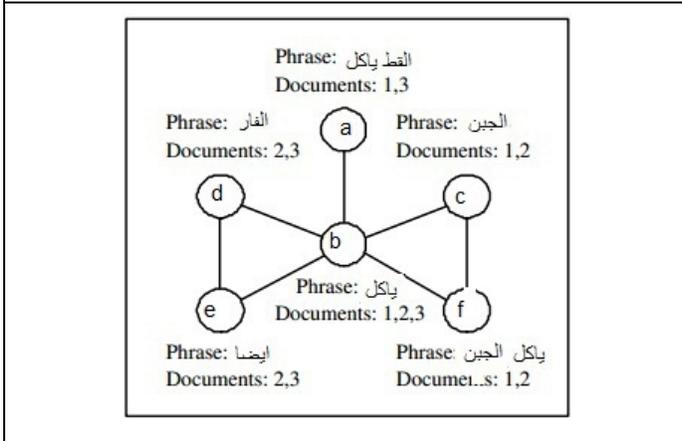

Figure.3. Base Cluster Graph for **Arabic language**

**English Documents:**

Document 1 : "cat ate cheese",                                    A

Document 2: "mouse ate cheese too",

Document 3: "cat ate mouse too",

```
STDM {                                                           B
      1 -> [label="cat ate" doc:(1,3)];
            2 -> [label="cheese $" doc:(1)];
            2 -> [label="mouse too $" doc:(3)];
      1 -> [label="ate" doc:(1,2,3)];
            2 -> [label="cheese" doc:(1,2)];
                  3 -> [label="$" doc:(1)];
                  3 -> [label="too $" doc:(2)];
            2 -> [label="mouse too $" doc:(3)];
      1 -> [label="cheese" doc:(1,2)];
            2 -> [label="$" doc:(1)];
            2 -> [label="too $" doc:(2)];
      1 -> [label="$" doc:(1,2,3)];
      1 -> [label="mouse" doc:(2,3)];
            2 -> [label="ate cheese too $" doc:(2)];
            2 -> [label="too $" doc:(3)];
      1 -> [label="too $" doc:(2,3)];
}
```

Output of our Java Progamme

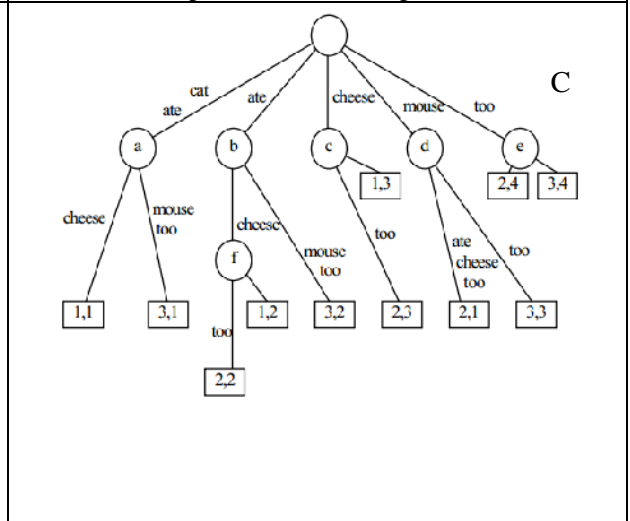

C

Figure.2. An instance of the Suffix Tree using **English Language** (A,B,C)

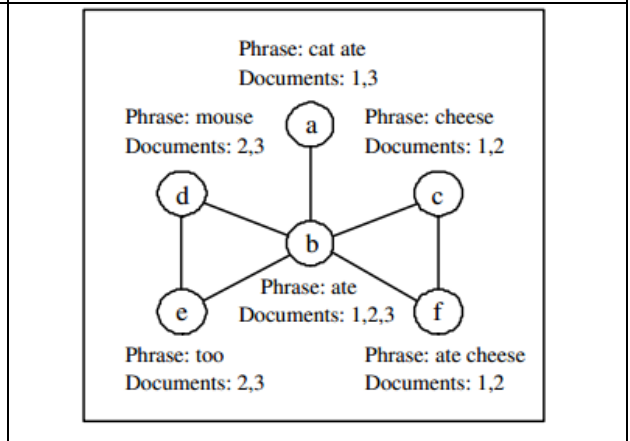

Figure.4. Base Cluster Graph for **English Language**

### 3.3. Suffix Tree Clustering for Arabic Language: Merging Process Problem

Suffix Tree Clustering (STC) algorithm has been introduced and used successfully and extensively with different adapted versions for English, European, and Chinese and other Languages. However, up to the present day, in our knowledge, STC algorithm has been never applied to cluster Arabic Web Search Results (**Snippets**). In this subsection, we illustrate by example the main problem related to Arabic Language pre-processing when using STC.

The generation of the **Suffix Tree** is based on comparison of each word in the Snippets with other all words in the **Snippets** corresponding to the obtained returned search results. In fact, if two words have two different sets of letters even if they have the same meaning or may be between two words one in singular form and other in plural form they will considered as two different words, and therefore generate two different nodes or clusters, this will influence negatively the merging process. To solve this problem for English language, Zamir et al. [2] have been apply stemming step using Porter algorithm [16 ] to extract a root for each words and use it for the comparison during the Suffix tree construction process. In our case, this solution is not efficient and not suitable for Arabic Language. As we had seen previously, Arabic Language is significantly different to Latin language and as the example (Figure.5 ) shows, that's when we have used the Stemming process before the Suffix Tree generation, it' will influence negatively the quality of clustering results: we have too much clusters that have no signification with user query (Figure.5 ) (Islam, ).

| Label cluster | English translation |
|---|---|
| node:ﺗﻲ doc:(0,15,9) | Drape |
| node:ﺧﺒﺮ doc:(1,3) | News |
| node:ﻗﺪﻡ doc:(2,5) | Offer |
| node:ﻧﺒﺄ doc:(2,12,3) | News |
| node:ﺑﯿﺾ doc:(2,9) | Whiten |
| node:ﺩﺭﺱ doc:(3,8) | Study |
| node:ﺯﻭﺭ doc:(6,11) | Tamper |
| node:ﺛﺒﺖ doc:(7,8) | Stand |
| node:ﺑﺪﺃ doc:(7,15) | Start |
| node:ﺑﻠﻎ doc:(8,15) | Reach |
| node:ﻛﺸﻒ doc:(9,15,0,18) | Detect |
| node:ﺭﺟﻊ doc:(12,13) | Return |
| node:ﺻﻠﻲ doc:(12,18) | Pray |
| node:ﺿﺨﻢ doc:(4,7,14,12,13) | Extend |
| node:ﺩﻋﺎ doc:(11,15,18,9) | Invite |
| node:ﻭﻗﻊ doc:(1,3,6,12) | Sign |
| node:ﺷﺮﻉ doc:(5,8,13,15) | Legislate |
| node:ﻇﮭﺮ doc:(0,3,4,5,13,14) | Know |
| node:ﺳﻠﻢ doc:(0,1,2,3,4,5,6,7,8,11,12,14,15,18) | Grant |

Figure .5. the obtained Clusters Results using STC for ( ,islam) Query after Arabic Root Extraction

The obtained results shown in (figure.5.) illustrate that when using STC after Arabic Root Extraction will influence negatively the quality of clustering results. To overcome this problem, we propose in the following section to integrate STC in a new scheme by applying STC clustering before Arabic Root Extraction.

# 4. SUFFIX TREE CLUSTERING FOR ARABIC LANGUAGE: PROPOSED NEW SCHEME

In this section, we present our proposition to solve the above problem related to application of STC algorithm for Arabic Language.

## 4.1. Flowchart:

Our proposed new scheme can be presented by the Flowchart (Figure.6) and summarized as follow: The Arabic user specify the Arabic query using the web interface, the query will be sent to the Google Web Search Engine using the services offered by the Google API [17]. The list of returned results are in the form of **Snippets** (id, link, body, title) figure.7, then we remove Arabic stop words, Latin words and specials characters like ( /, #, $, ect…). After, each Snippet is used to generate the Suffix Tree (Figure.4). The Suffix Tree is used to find all **Base Clusters**. These latter are scored and the highest scoring is merged into clusters to construct the **Base Clusters Graph** (Figure.3). Finally, we propose to find the root of each label node using Khoja stemmer [18] in order to merge all similar clusters and eliminate all no significant clusters.

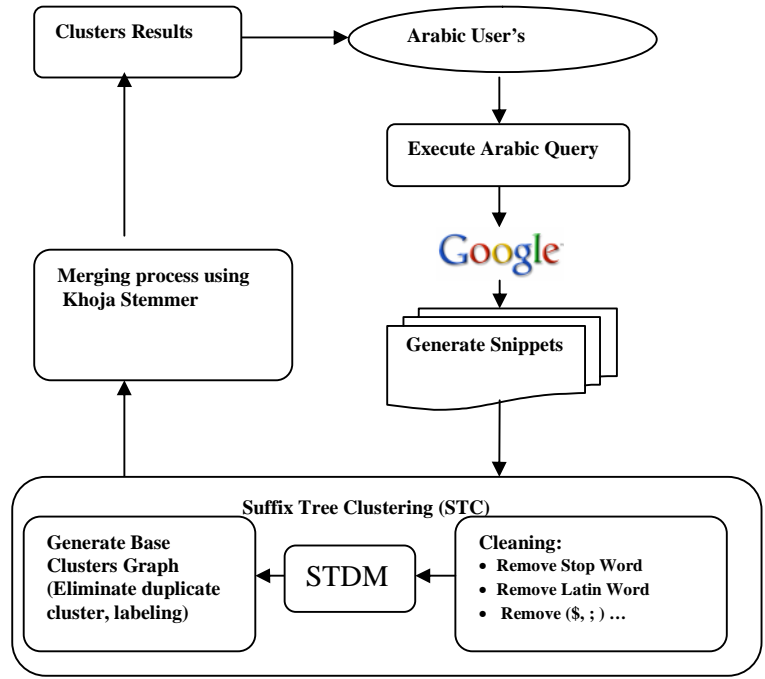

Figure.6: Our proposed new scheme for STC applied to Arabic Language.

```
<snippet>
  <id>0</id>
  <url>http://ar.wikipedia.org/wiki/%D8%AA%D8%89%D9%84%D9%8A%D9%85</url>
  <body>... التطم والتربية هو بناء الفرد وهو المجتمع. وهو المحركة الأساس في تطور  الحضارات ومحور قياس تطور وتقدم المجتمعات فتقم تلك المجتمعات على حسب بـ...</body>
  <title>تطم - ويكمبدا: الموسوعة الحرة</title>
</snippet>
<snippet>
  <id>1</id>
  <url>http://www.mohe.gov.sa/</url>
  <body>المملكة العربية السعودية - يحمل الموقع كل ما بحاجة المتصفح من معلومات خاصة  بالتطم العالي بالمملكة.</body>
  <title>وزارة التطم العالي</title>
</snippet>
```

Figure.7. An example of two Arabic Web **Documents Snippets**

| Cluster Labels | English translation |
|---|---|
| node: والعلوم doc:(4,14,7) | Science |
| node: الإسلامية doc:(8,12,3,5,7,0,2,4,6,11,14) | Islamic |
| node: المراجع doc:(12,13) | Reference |
| node: موقع doc:(1,3,6,12) | Location |

Figure 8.Clusters Results using STC  for (    ,islam) Query before Arabic Root Extraction

The obtained results in (Figure.8) shows that our proposition for Arabic **Snippets** clustering presents good results in comparison with those presented in (Figure.5). In the following section, we present our new system for Arabic Web Search Results Clustering using Google API search engine.

# 5. EXPERIMENT RESULTS

## 5.1. Our Proposed Platform for Arabic Web Search Results Clustering AWSRC

The implement of our clustering search engine based on STC algorithm contains five modules:
1. **Query module**: it provides the input Web Interface for Arabic user (Figure.9).

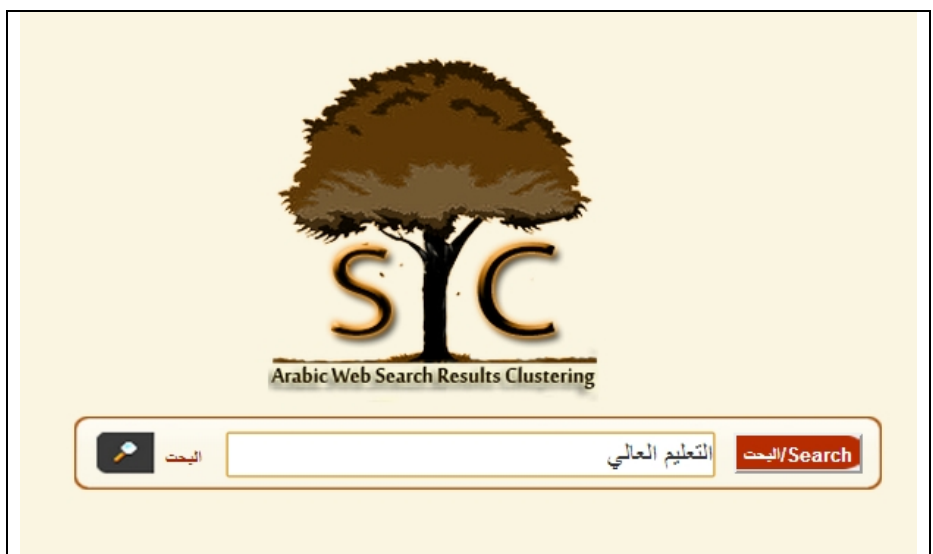

Figure.9. AWSRC platform Interface (Arabic Query example: التعليم العالي / High Education).

26

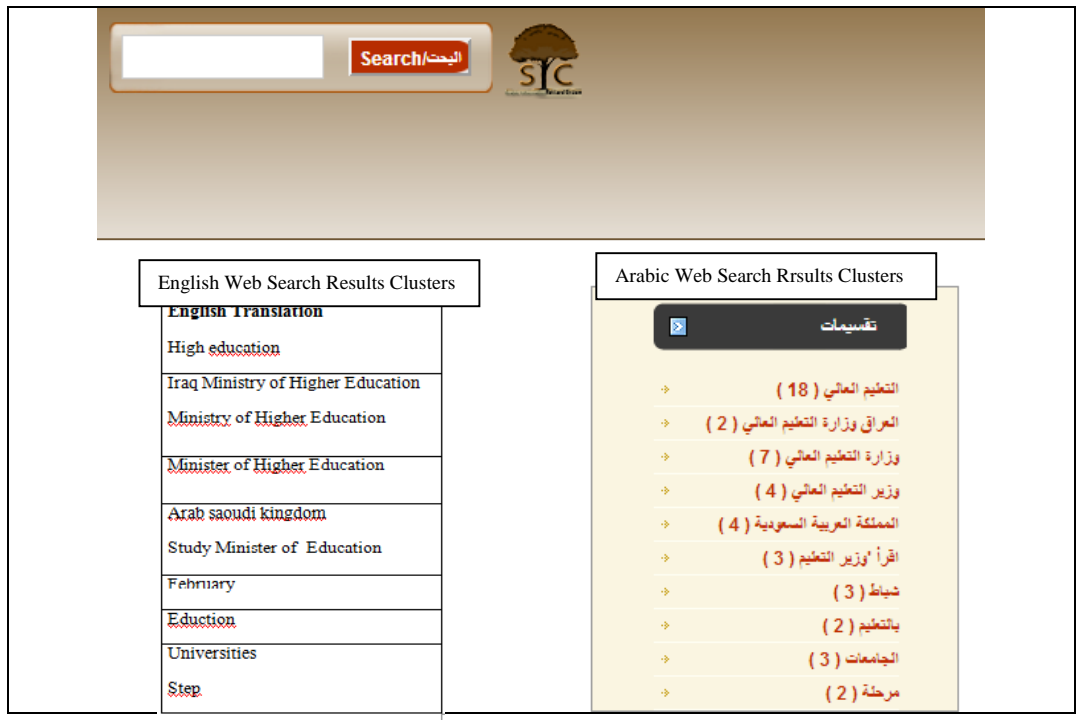

Figure.10. AWSRC platform Interface (High Education/التعليم العالي)

When a user fills in keywords in the query interface, the query will be sent to the interface of Google Web API, namely, the query keywords are sent to Google search engine for querying. These search results will be returned to analyze module as the form of snippets and save in XML Documents.

2. **Analyze Module**: it is used to finish pre-processing for result search, including deleting repeat snippets, generate clustering document.

3. **Clustering Module**: build suffix tree based on STC algorithm by the input of clustering document from step 2 and save nodes whose cluster labels meet the requirement by accessing nodes in tree to result linklist.

5. **Use the khoja Stemmer** after de STC algorithm we can overlapping all similar clusters and eliminate all additional clusters.

4. **Result Display Module**: the page which is used to display results contains two parts. One part in the right denotes a tree structure for cluster label, the other part in the left denotes the content which belonging to (Figure.10).

## 5.2. Experiment Results

Generally, the evaluation of traditional clustering techniques may be divided into two approaches: the objective approach and the subjective one. Note that, in the objective approach, original partition is known and compared to generated clusters. The comparison may be performed using traditional measures from clustering evaluation domain such as Entropy and Purity used in our previous works [20][21]. However, objective evaluation of Search Results Clustering SRC algorithms is very difficult; in fact there is no standard test collection for evaluating Search Results Clustering SRC algorithms, because we have any prior knowledge about the initial clusters corresponding to user's queries. Therefore, in this paper, we propose to use only the subjective approach.

To evaluate and illustrate the effectiveness of our proposed algorithm, we present and discuss some top result of some Arabic Query tab.1. Our obtained results are also compared with the three others existing Search Web Results Clustering System such as **Clusty**, **IBoogie**, **Yippy**.

Table.1 Some Arabic query

| Query | التعليم/**Education** | السيارات/**cars** | السياحة/**tourism** |
|---|---|---|---|
| Results | التعليم/Eduction | اخبار السيارات/ Cars News | السياحة / tourism |
| | وزير التعليم/ Minister of Education | العاب سيارات/ Cars Games | السياحة نشاط / tourism activity |
| | التربية والتعليم/Eduction | سباقات السيارات/ Race Cars | travel/ |
| | Center / | Last News / | services / |
| | Build / | Race Games / | Egypt / |

This table presents five top results of some Arabic queries. As we observe that the most of this cluster label of each query are Keywords belongs to the same domain. In the following section we will search about "السياحة, tourism" using the three famous web post-retrieval System **Clusty** (Figure.10) , **IBoogie** (Figure.11) and **Yippy** (Figure.13) (formerly **Clusty** [19] ), in Order to compare their results with the obtained results using our proposed system (AWSRC) (Figure.12). From (Figure.10), (Figure.11), (Figure.12) and (Figure.13) we can conclude the following (Table.2).

In Table.2, we have base on two most important factors to evaluate system with Arabic language:

1. Filtering to display just Arabic results
2. How Arabic language have been treated

Table.2 Comparative study of Post Information Retrieval System

| | Clusty | IBoogie | Yippy | AIRS |
|---|---|---|---|---|
| *How Arabic language have been treated?* | Use cross language | Use cross language | No Arabic language | Arabic Language |
| *Filtering to display just Arabic results* | Mixture of language Results | Latin language Results | No Arabic Results | Arabic results |

From Table.2, we can conclude that our system is efficient and the most suitable for Arabic language than the others existing Search Web Results Clustering System.

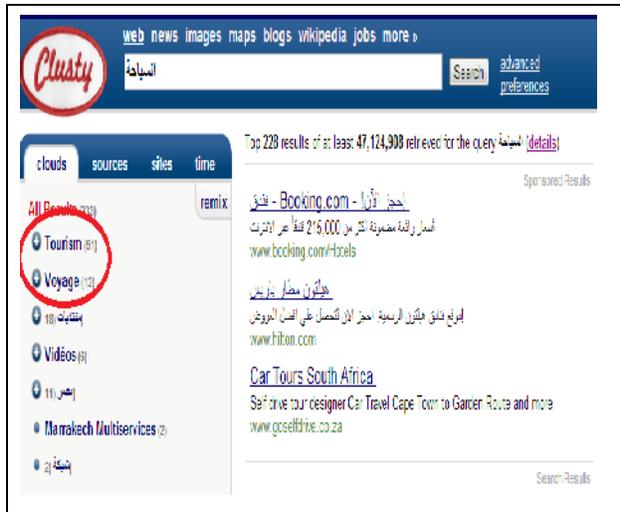

Figure..10.  Clusty  System Diplay Results (tourism/السياحة)

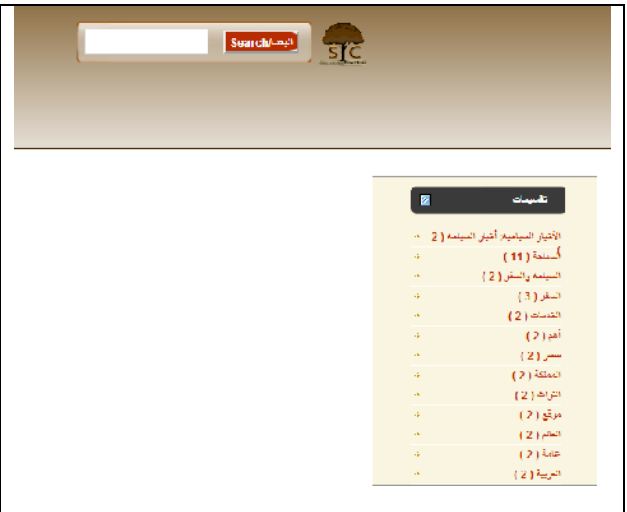

Figure.12.   AWSRC  System Diplay Results (tourism/السياحة)

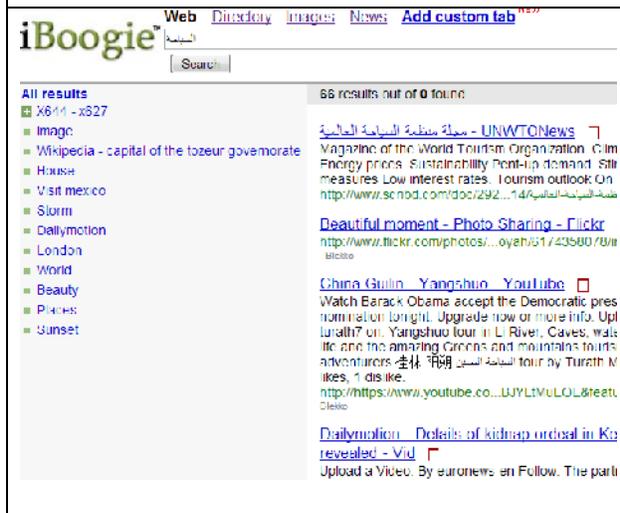

Figure .11. iboogie  System Display Results (tourism/السياحة)

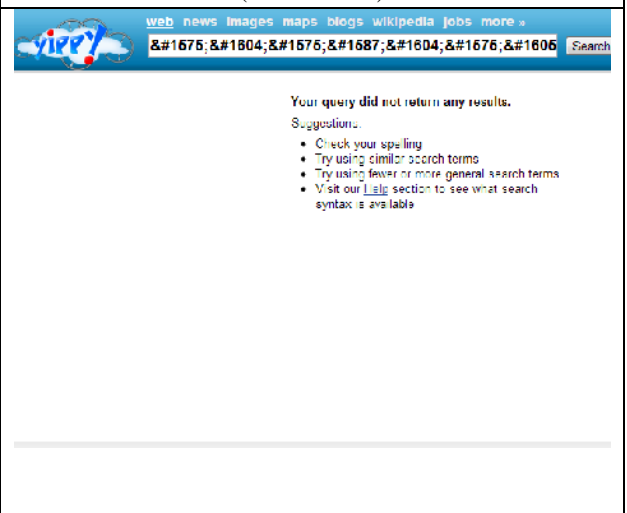

Figure .13. yippy   System Display Results (tourism/السياحة)

# CONCLUSION

Browsing Search Results is one of the major problems with traditional Web search engines (Google, Yahoo and Bing) for English, European, and any other languages generally and for Arabic Language particularly. Organizing Arabic web search results into clusters facilitates Arabic users quick browsing through web search results.

Suffix Tree Clustering (STC) algorithm has been successfully and extensively used with different adapted versions for English, European, and Chinese and other Languages. However, till the day of writing this paper, in our knowledge, STC algorithm has been never applied for Arabic Web Snippets Search Results Clustering.

In this paper, we have illustrate that the STC algorithm can't be applied directly for Arabic Language, after we have proposed, and implemented a new scheme that integrate correctly the STC algorithm in our framework for Arabic Web Search Results Clustering. The proposed approach automatically clusters the web search results into high quality, and high significant clusters labels. Preliminary experiments and evaluations are conducted and the experimental results show that our proposition is effective and promising to facilitate Arabic users quick browsing through search results. Finally, a recommended platform for Arabic Web search results clustering is established based on Google search engine API.

In our future work, we suggest integrating and using the Yahoo API, Bing API and evaluate the proposed platform with some no professional web user's to calculate satisfaction factor.

**Authors**

**Issam SAHMOUDI** received his Computer Engineering degree from "Ecole Nationale des Sciences Apliquées, FEZ (ENSA-FEZ). Now he is PhD Student in Laboratory of Information Science and Systems LSIS, at (E.N.S.A), Sidi Mohamed Ben Abdellah University (USMBA), Fez, Morocco. His current research interests include Arabic Text Mining Applications: Arabic Web Document Clustering and Browsing, Arabic Information and Retrieval Systems, and Arabic web Mining.

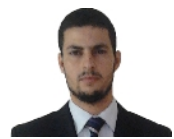

**Pr. Abdelmonaime LACHKAR** received his PhD degree from the USMBA, Morocco in 2004, He is Professor and Computer Engineering Program Coordinator at (E.N.S.A), and the Head of the Systems Architecture and Multimedia Team (LSIS Laboratory) at Sidi Mohamed Ben Abdellah University (USMBA), Fez, Morocco. His current research interests include Arabic Natural Langage Processing ANLP, Arabic Text Mining Applications: Arabic Web Document Clustering and Categorization. Arabic Information and Retrieval Systems, Arabic Text Summarization, and Arabic web Mining.

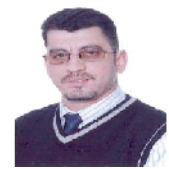